\def\k{{\bf k}\/}
\def\be{\begin{equation}}
\def\ee{\end{equation}}
\def\ber{\begin{eqnarray}}
\def\eer{\end{eqnarray}}
\def\bers{\begin{eqnarray*}}
\def\eers{\end{eqnarray*}}

\def\PR{{ Phys. Rev.}\ }
\def\PRL{{ Phys. Rev. Lett.}\ }

\documentclass[aps,prb,twocolumn,groupedaddress,showpacs,amsmath,amssymb]{revtex4-1}
\usepackage{dcolumn}   
\usepackage{amsmath}
\usepackage{graphicx}    
\usepackage{subfigure}
\usepackage{color}
\newcommand{\comment}[1]{}
\usepackage{ifthen}
\newboolean{includefigs}
\setboolean{includefigs}{true}      
\newboolean{includetext}
\setboolean{includetext}{true}     
\newcommand{\condcomment}[2]{\ifthenelse{#1}{#2}{}}
\begin{document}
\voffset 2cm
\title{Thermal conductivity and diffusion-mediated localization in Fe$_{1-x}$Cr$_{x}$ Alloys}

\author{Aftab Alam}
\email[emails: ]{aftab@ameslab.gov,  abhijit@bose.res.in}
\affiliation{Division of Materials Science and Engineering, Ames Laboratory, Ames, Iowa 50011, USA}
\author{Rajiv K. Chouhan$^{1}$, and Abhijit Mookerjee$^{2}$}
\affiliation{$^1$ Department of Materials Science, S.N. Bose National Centre for Basic Sciences, JD-III Salt Lake City, Kolkata 700098} 
\affiliation{$^{2}$Advanced Materials Research Unit and Department of Materials Science, S.N. Bose National Centre for Basic Sciences, JD-III Salt Lake City, Kolkata 700098}

\begin{abstract} 
We apply a new Kubo-Greenwood type formula combined with a generalized Feynman diagrammatic technique to report a first principles calculation of the thermal transport properties of disordered Fe$_{1-x}$Cr$_{x}$ alloys. The diagrammatic approach simplifies the inclusion of disorder-induced scattering effects on the two particle correlation functions and hence renormalizes the heat current operator to calculate configuration averaged lattice thermal conductivity and diffusivity. The thermal conductivity $\kappa(T)$ in the present case shows an approximate quadratic $T$-dependence in the low temperature regime ($T<20$ K), which subsequently rises smoothly to a $T$-independent saturated value at high $T$. A numerical estimate of mobility edge from the thermal diffusivity data yields the fraction of localized states. It is concluded that the complex disorder scattering processes, in force-constant dominated disorder alloys such as Fe-Cr, tend to localize the vibrational modes quite significantly.
\end{abstract} 
\date{\today}
\maketitle

     
 The study of phonon excitations and the associated thermal transport properties is an important field of research in disordered alloys.
 In certain materials, disorder mediated scattering can shrink the typical mean free path (MFP) of phonons to such a level that wavelength and MFP no
 longer remain sharp concepts, and the usual textbook phonon gas  model for thermal conductivity breaks down. From the theoretical perspective, the 
development of a reliable quantum mechanical theory to predict such properties in random alloys is a difficult task mainly because of two problems :
 (i) one needs a microscopic description of inter-atomic force constants with an intrinsic off-diagonal disorder and (ii) one has to configuration average
 a two-particle correlation function within a Kubo-type formula. The effects of dominant off-diagonal force constant disorder in alloys can be quite unusual, 
as we have shown earlier.\cite{Alam1} Most theories of thermal transport, developed in the past few decades however, are either based on the single-site
 coherent potential approximation (CPA),\cite{CPA} the perturbation-based approach simulating the Peierls-Boltzmann equation (PBE)\cite{Sun2010} or
 atomistic models with a large unit cell and periodic boundary conditions.\cite{Allen93} CPA, being a single-site mean-field approximation, is inadequate 
for treating multi-site off-diagonal disorder arising out of force constants and is reported, for example, to inadequately explain experimental life-time data on simple Ni-Pt alloys.
\cite{Tsunoda79} The perturbative simulation approach, although rigorously derived, is limited in applicability to model lattices alone and 
has  not been tested on realistic materials. The atomistic models are computationally expensive due to the large unit-cell size, non-self-consistent and suffer from the finite size errors.

  In a recent paper\cite{Alam2} we have developed a theoretical approach to calculate the configuration averaged lattice 
thermal conductivity and diffusivity
 for random alloys. This formalism combined a Kubo-Greenwood approach with a generalized Feynman diagrammatic technique to 
explicitly incorporate the effect of
 disorder induced scattering.  We showed that disorder scattering  renormalizes both the phonon propagators as well as the heat currents. These corrections 
 are related to  the self-energy and vertex corrections. Unlike the single-site CPA, this approach explicitly takes into account
 the fluctuations in masses (diagonal)
, force constants and heat currents (off-diagonal disorder) between different ion-cores and incorporates the sum rule relating 
the diagonal element of the force constant
to the off-diagonal ones. 

In the present paper, we combine this theoretical approach with a first-principles Quantum-ESPRESSO (QE) calculation\cite{QESPRESSO}
 of the force-constants. QE is a linear response based method : the  density functional perturbation theory (DFPT).\cite{DFPT}
 The dynamical matrix for the phonon excitation of a system is obtained from the ground state electron charge density and its linear response to a 
distortion of the ion-core geometry. We refer the reader to a recent article\cite{Alam3} for further  computational details on Fe$_{1-x}$Cr$_x$ alloys.
This alloy, being a basic ingredient of stainless steel, is a technologically important structural material, dominated by  force-constant disorder and hence should serve as a critical test of our theory for the thermal transport properties. 

 We find that the disorder induced scattering effects on the thermal conductivity, $\kappa(T)$, is relatively large in the low
 frequency regime.  $\kappa(T)$ shows a quadratic $T$-dependence in the low temperature range, where only low
 energy vibrations are excited, and then smoothly rises to a $T$-independent saturated value at high $T$. Thermal diffusivity
 manifests the effect of disorder in a more dramatic fashion, and gives an idea about localization. Based on our calculation  on Fe$_{1-x}$Cr$_x$ alloys, a large fraction ($>$ 90~\%) of vibrational eigenstates are found to be localized with the maximum
 localization near 50-50 composition, where the disorder scattering is maximum, as expected.

  For disordered materials, the lattice thermal conductivity requires the configuration average of the response functions of the kind (see Ref. \onlinecite{Alam2}),
\begin{eqnarray}
\langle\langle \kappa(z_1,z_2,T) \rangle\rangle & =&\nonumber\\
&&\hspace{-2.5cm}\int \frac{d^3{\bf k}}{8\pi^3} \text{Tr} \left[\rule{0mm}{3mm}\langle\langle {\bf S}({\bf k},T) {\bf G}({\bf k},z_1) {\bf S}({\bf k},T) {\bf G}({\bf k},z_2) \rangle\rangle\right], 
\label{eq1}
\end{eqnarray}
where ${\bf S}$ is the heat current operator and ${\bf G}$ is the phonon propagator.$\langle\langle\ \ \rangle\rangle$ denotes configuration averaging.

\begin{figure}[t]
\centering
\includegraphics[width=8.5cm]{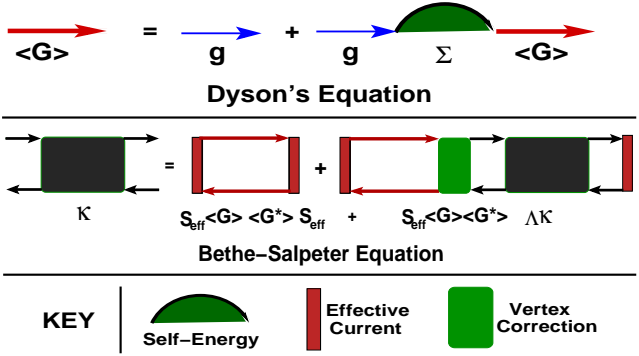}
\caption {(Color Online) (Top row) Dyson's Equation due to scattering diagrams for the single particle averaged Green's
functions for disordered alloys. (Middle Row) Bethe-Salpeter equation for the response functions in disordered alloys. (Bottom Row) Key to diagrams. 
$<{\bf G}>$ is the averaged disorder renormalized Green's function, $<\kappa>$ is the two-particle correlation function related to  Thermal conductivity, $\Sigma$ is the self energy and {\bf S}$_\text{eff}$ is the disorder-renormalized effective current.}
\label{fig1}
\end{figure}

 The right hand side of Eq. \eqref{eq1} involves the configuration average of four random functions whose 
fluctuations are correlated. Unlike the configuration average of  a single 
particle Green function $\langle\langle {\bf G}({\bf k},z)\rangle\rangle$, which can be calculated via a 
perturbative self-consistent Dyson's equation (shown diagrammatically in the
 1st row of Fig. \ref{fig1}), the average of a two-particle correlation
 function such as that in Eq. \eqref{eq1} is non-trivial. The zeroth order
 approximation for such an average is the one which assumes the fluctuations
 between all four random functions to be uncorrelated, and expresses the average
 of the product as the product of the averages (as in the so called Virtual Crystal Approximation (VCA)). 
 The inherent correlation, however, requires the contributions from averages
 taken in pairs, triplets and all four random functions. Such disorder induced
 corrections can be calculated very efficiently within a Feynman diagrammatic 
technique (details have been discussed in Ref. \onlinecite{Alam2}), which renormalizes both
 the phonon propagators as well as the heat currents to provide a mathematical
 expression for $\langle\langle\kappa\rangle\rangle$ with an {\it effective} heat current ${\bf S}_{\text{eff}}$
 related to the self-energy of the propagators (shown by 1st diagram on RHS of the middle row of Fig. \ref{fig1}). The last term in the middle row gives the contribution from
 the so called vertex correction arising out of the correlated propagation.
 For a harmonic solid, thermal diffusivity has a similar expression as $\langle\langle\kappa\rangle\rangle$
 except the product of five random functions instead of four.
 A similar diagrammatic procedure has been used earlier by us \cite{Alam2} to calculate the configuration 
averaged thermal diffusivity as given by,
\begin{eqnarray}
\langle\langle D(\nu)\rangle\rangle &=& 
\frac{1}{\pi^2}\int d\nu' \int \frac{d^3{\bf k}}{8\pi^3}\nonumber\\
&&\hspace{-1.2cm} \text{Tr}\left[\rule{0mm}{3mm} \langle\langle\Im m {\bf G}(\k,\nu')
{\bf S}(\k) \Im m {\bf G}(\k,\nu') {\bf S}(\k) \Im m {\bf G}(\k,\nu)\rangle\rangle \right].\nonumber
\end{eqnarray}

\begin{figure}[t]
\centering
\includegraphics[width=7.5cm]{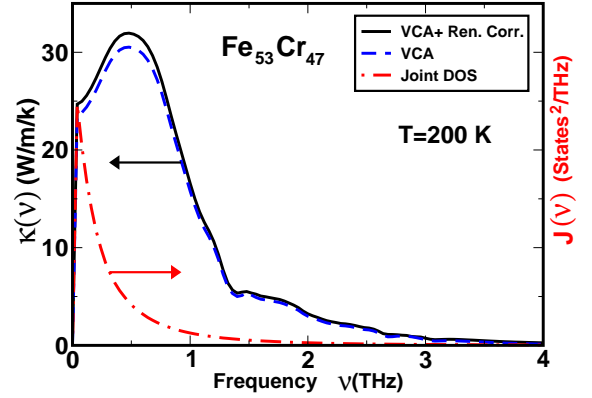}
\caption {(Color Online) Frequency dependence of Thermal conductivity and Joint density of states  for Fe$_{53}$Cr$_{47}$ alloy at $T=200$ K. Solid line shows the result including all disorder-induced corrections + the vertex correction (middle row of Fig. \ref{fig1}) and dashed line including the VCA average alone.}
\label{fig2}
\end{figure}

{\par} In Fig. \ref{fig2}, we display the frequency dependence of lattice
 thermal conductivity $\kappa(\nu)$ and the scaled joint density of states
 $J(\nu)$ at $T=200$~K for the Fe$_{53}$Cr$_{47}$ alloy. It is obvious from the
figure that the transition rate $\tau$ (related to the heat current operator)
is strongly dependent both on the initial and final energies throughout the 
phonon spectrum i.e. $\kappa(\nu,T) \ne \vert\tau(\nu,T)\vert J(\nu)$, 
where $J(\nu)$ (shown by dot-dashed line in Fig. \ref{fig2}) is given by

\begin{eqnarray*}
\langle\langle J(\nu) \rangle\rangle &=\\
&& \hspace{-1.5cm}\int d\nu'\int \frac{d^3{\bf k}}{8\pi^3} \text{Tr} \left[\rule{0mm}{3mm} \Im m\langle\langle{\bf G}({\bf k},\nu')\rangle\rangle \Im m\langle\langle{\bf G}({\bf k},\nu'+\nu)\rangle\rangle\right].
\end{eqnarray*}

The effect of disorder-induced renormalized corrections (black solid lines) to the zeroth order virtual-crystal-approximation (VCA)
 (blue dashed line) thermal conductivity is not significant, and becomes negligibly small beyond $\nu=2.7$~THz. The traditional single site mean-field approximation is, therefore, expected to describe well the multiple scattering phenomenon associated with the high frequency mode, deviating only in the low frequency range where the higher order corrections become important. Notably, both 
$\kappa(\nu)$ and $J(\nu)$ curve has a dip at a very small energy ($\nu\simeq 0$). Such a dip reflects the missing intraband contribution ($\kappa^{II}$) to the 
conductivity. The origin of this dip is a natural 
outcome of a smooth convolution of two Green matrices ${\bf G} ({\bf k},\nu')$
 and ${\bf G} ({\bf k},\nu'+\nu)$ (or two smooth DOS).  
 A similar dip has  also been reported by Feldman 
{\emph {et al.}}\cite{Allen93} in amorphous Si and Si$_{1-x}$Ge$_x$ alloys.
Unlike our case ($\kappa(\nu)\rightarrow 0$ as $\nu\rightarrow 0$), this dip
in their calculation stands at a finite value as $\nu\rightarrow 0$.
These authors have introduced an arbitrary   
 Lorenzian broadening of the delta
functions in their Kubo-Greenwood expression for $\kappa$, while in our 
calculation this arises naturally through the disorder induced broadening of the 
spectral function $\Im m[{\bf G} ({\bf k},\nu)]$. An extrapolation of our
$\kappa(\nu)$-curve (see Fig. \ref{fig2}) from a value just above $\nu=0$ to 
a value at $\nu=0$ yields an estimate of the dc thermal conductivity, which
 comes out to be $24.7$~W/m/K for the present Fe$_{53}$Cr$_{47}$  alloy 
at $T=200$~K. 
Literature survey shows a lack of available experimental data for concentrated 
Fe$_{1-x}$Cr$_{x}$ alloy, however there exist some on the dilute Cr-alloys.
\cite{new} For example $\kappa_{\text{expt}}$ for $x=0.25\%$ Cr is $\sim$  
$22$~W/m/K, with which we shall compare our theoretical estimate below. 

\begin{figure}[t]
\centering
\includegraphics[width=7.5cm]{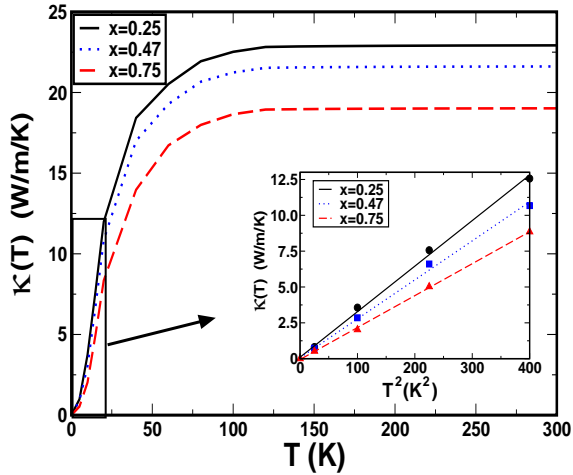}
\caption {(Color Online) Temperature dependence of Thermal conductivity ($\kappa$) for three Fe$_{1-x}$Cr$_x$ alloys. Inset shows the quadratic $T$-dependence of $\kappa$ in the low $T$-range. }
\label{fig3}
\end{figure}

\begin{figure}[t]
\centering
\includegraphics[width=8cm]{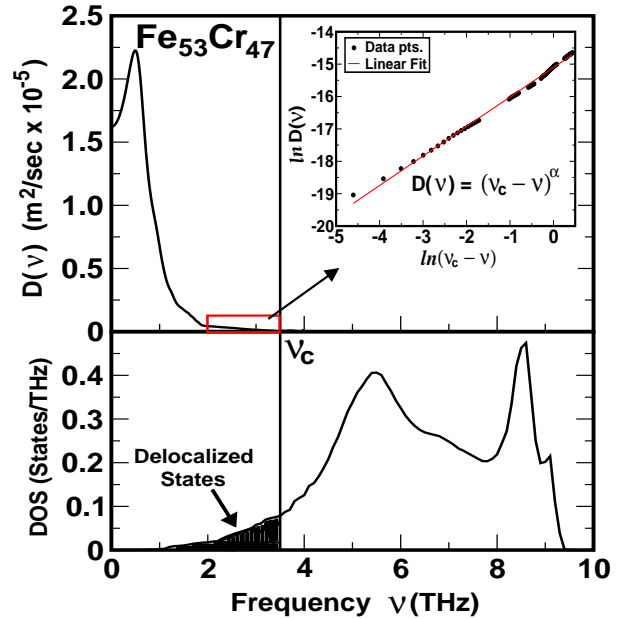}
\caption {(Color Online) Thermal diffusivity (top) and DOS (bottom) vs. phonon frequency ($\nu$) for Fe$_{53}$Cr$_{47}$ alloy. Inset shows an approximate linear $\nu$-dependence of $D$($\nu$) above $2$~THz. $\nu_c$ locates the mobility edge and area under the shaded region gives an estimate of the fraction of delocalized states. }
\label{fig4}
\end{figure}

{\par} Figure \ref{fig3} shows the temperature dependence of thermal conductivity for
three Fe$_{1-x}$Cr$_x$ alloys. Note that $\kappa(T)$ behaves quadratically 
(see inset) in the low temperature regime ($T<20$~K) where only low-energy 
vibrations are excited. As the temperature is increased further, the 
$T$-dependence of $\kappa$ becomes much milder and eventually reaches a 
$T$-independent saturated value. The origin of such a high $T$-saturation is not
very well described by most previous theories. 
Within a harmonic approximation,
 such a saturation mainly arise from the $T$-dependence of Einstein specific
 heat piece of the conductivity expression.\cite{Allen93} The intrinsic harmonic diffusion
 of higher energy delocalized vibrations are mostly responsible for the
 relevant dominant mechanism in this regime. Another qualitative explanation can
be that : the phonon-phonon scattering in this high $T$-range becomes so strong
that the phonon MFPs reach a minima, and further enhancing the 
disorder scattering by raising temperature would not cause any further 
reduction in the MFP, hence resulting in a $T$-independent thermal conductivity. This, however, is just a physically plausible explanation based on the MFP and
 is not intended to reflect a known outcome of the proposed theory itself.
One can also notice an overall reduction of $\kappa$ with 
increasing disorder $(x)$, as expected. Such effect usually reflect the  
 scattering arising out of the difference in masses, radii and
force constants between the host lattice atoms and impurities. In the present
Fe$_{1-x}$Cr$_x$ alloy, however, this scattering is mainly dominated by a
large difference of force constants between Fe and Cr atoms in the alloy, while 
their masses and radii are almost similar. 
As far as the comparison goes, looking at the solid black curve of Fig. \ref{fig3} for x=0.25\% Cr, one can notice the saturated value of thermal conductivity (room temperature value) to be $22.8$~W/m/K which is in good agreement with the experimental value of $22$~W/m/K.

{\par} Next, we examine the effect of disorder scattering on the vibrational 
eigenstates and hence the localization of the phonon modes based on a thermal 
diffusivity calculation. 
In Fig. \ref{fig4}, we show the thermal diffusivity
(top panel) and the phonon density of states (bottom panel) vs frequency 
for the Fe$_{53}$Cr$_{47}$ alloy. Above $\nu\simeq2$~THz, D($\nu$) decreases
smoothly (approximately linear in $\nu$) with a critical frequency 
$\nu_c = 3.55$~THz, where D($\nu$) vanishes to within a very small level of
noise. This regime is shown, for clarity, as a log-log plot within the inset of 
the top panel. The calculated critical exponent $\alpha\simeq1.011$ agrees 
with the scaling and other theories of Anderson localization.\cite{Lee85} The
critical frequency $\nu_c$ locates the mobility edge above which the 
diffusivity strictly goes to zero in the infinite size limit, and the allowed
vibrational states above (below) this frequency remain localized (delocalized).
 This is shown by the area under the shaded (unshaded) region in the bottom 
panel which gives an estimate of the  percentage of delocalized (localized) 
states.

\begin{figure}[t]
\centering
\includegraphics[width=6.5cm]{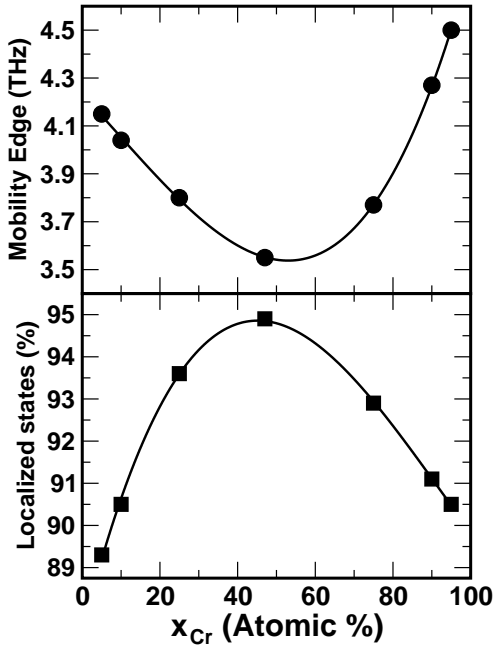}
\caption {Mobility edge (top) and the percentage of localized states (bottom) vs Cr-concentration for Fe$_{1-x}$Cr$_x$ alloy.}
\label{fig5}
\end{figure}

The  location of mobility edge ($\nu_c$) and the percentage of localized
states (calculated from the area under the curve in Fig. \ref{fig4}) with varying Cr-content in Fe$_{1-x}$Cr$_x$ alloy is shown 
in Fig. \ref{fig5}. Such a non-monotonous variation of the fraction of localized states is an 
artifact of the varying band-width of the phonon spectrum with $x$. Unlike 
the mass dominated Si$_{1-x}$Ge$_x$ alloys\cite{Allen93} which show an 
increasing percentage of localized states, towards the upper end of the phonon
 spectrum, with increasing Ge-concentration, the Fe-Cr alloys show maximum
localization at $x_{\text{Cr}}=47~\%$.
Such a maximal localization may be due to the dominance of the force constant
disorder in the present alloy which causes an enhanced disorder scattering at
$x=47~\%$ and hence localize the vibrational modes maximally.

 An alternative way of estimating the fraction of localized 
(delocalized) states is to calculate the so called ``{\it inverse 
participation ratio}'' $1/p_{\gamma}$ defined as, $1/p_{\gamma}=\sum_{\mu}\int (d^3k/8\pi^3)\ \epsilon_{\gamma}^{\mu} ({\bf k})$, where 
$\epsilon_{\gamma}^{\mu} ({\bf k})$ is the $\mu$th Cartesian component of the
normalized polarization vector of the $\gamma$th mode. $p_{\gamma}$ measures
the number of atoms on which $\gamma$th vibrational mode has significant
amplitude. $1/p_{\gamma}\rightarrow 0$ for delocalized mode, but remains
finite for localized modes. Although this procedure provides a quick assessment
of localization, it suffers from a shortcoming which arises quite often from  
the existence of an unexpected {\it few} localized modes in the low frequency 
regime as discussed earlier.\cite{Allen93,Biswas88} 
This is mainly due to a 
sensitive dependence of $p_{\gamma}$ on the boundary condition used in the 
concerned model. In other words, finite-size theory (even for large model 
systems) cause an unphysical gap at the bottom of the spectrum, and the states 
closest to this gap may be or appear to be localized. The same states in a 
macroscopic sample, however, may not be localized but propagating (or may be 
resonant).\cite{Allen93,Biswas88} The percentage of localized (de-localized) 
states calculated using the area above (below) $\nu_c$ of DOS curve and using
 p$_{\gamma}$ may differ, if calculated from such finite-size theories. 
However, being a k-space based formulation, our theory does not suffer from 
such differences and is free from the unexpected errors arising from the 
existence of few localized modes in the low energy regime. 

{\par} In summary, we combine a generalized Kubo-Greenwood type formula  
with the linear-response based QE calculation to make a first
principles prediction of the thermal conductivity and diffusivity
 of disordered Fe-Cr alloys. The effect of disorder-induced scattering on 
$\kappa$ is found to decrease with increasing phonon energy. Thermal conductivity shows
a quadratic $T$-dependence in the low $T$-regime, increasing smoothly to a
$T$-independent saturated value at high $T$. Thermal diffusivity provides an 
estimate of the location of mobility edge, which subsequently gives an idea 
 about the disorder-induced localization in the system. Vibrational modes in the
present Fe$_{1-x}$Cr$_{x}$ alloy are maximally localized at $x=47~\%$, where
the effect of disorder scattering is maximum.

AA acknowledges support from the U.S. Department of Energy BES/Materials Science and Engineering Division from contracts DEFG02-03ER46026 and Ames Laboratory (DE-AC02-07CH11358), operated by Iowa State University. This work was done under the Hydra Collaboration

\vspace{-0.5cm}

\end{document}